\documentclass[prb,aps,twocolumn,showpacs]{revtex4-1}%
\usepackage{graphicx,dcolumn,amsmath,color,longtable}
\usepackage{amsmath}
\usepackage{amsfonts}
\usepackage{amssymb}
\usepackage{multirow}
\usepackage{graphicx}%
\newcommand{\be}{\begin{equation}}
\newcommand{\ee}{\end{equation}}
\newcommand{\bea}{\begin{eqnarray}}
\newcommand{\eea}{\end{eqnarray}}

\begin{document}
\title{Higher-order contributions to the Rashba-Bychkov effect with application to Bi/Ag(111) surface alloy}
\author{Sz.~Vajna$^{1}$}
\author{E.~Simon$^{1,2}$}
\author{A.~Szilva$^{1}$}
\author{K.~Palotas$^{1}$}
\author{B.~Ujfalussy$^{3}$}
\author{L.~Szunyogh$^{1}$}
\email{szunyogh@phy.bme.hu}
\affiliation{
$^{1}$Department of Theoretical Physics, Budapest University of Technology and
Economics, Budafoki \'ut 8., H-1111 Budapest, Hungary \\
$^{2}$L\'or\'and E\"otv\"os University, Department of Physics, H-1518 Budapest POB 32, Hungary \\
$^{3}$Hungarian Academy of Sciences, Institute for Solid State Physics and Optics, H-1525 Budapest, PO Box 49 H-1525 Hungary
}
\date{\today}

\begin{abstract}
In order to explain the anisotropic Rashba-Bychkov effect observed in several metallic surface-state systems,
we use $k \cdot p$ perturbation theory with a simple group-theoretical analysis and construct
effective Rashba Hamiltonians for different point groups up to third order in the wavenumber.
We perform relativistic \emph{ab initio} calculations for the $(\sqrt{3}\times\sqrt{3})R30^{\circ}$ Bi/Ag(111)
surface alloy and from the calculated splitting of the band dispersion we find evidence of the predicted third-order terms.
Furthermore, we derive expressions for the corresponding third-order Rashba parameters
to provide a simple explanation to the qualitative difference concerning the Rashba-Bychkov splitting
of the surface states at Au(111) and Bi/Ag(111).
\end{abstract}
\maketitle

\section{Introduction}
Since the first experimental verification by LaShell \emph{et al.}~\cite{au111_LaShell-prl96}
the spin--orbit induced splitting of Shockley states on metallic surfaces called the Rashba-Bychkov
(RB) effect~\cite{Bychkov-Rashba-jpcm84} became into the focus of experimental and theoretical research.
These investigations range from the prototypical $L$-gap surface states at Au(111) and Ag(111)~\cite{auag111_Nicolay-prb01, au111_Henk-prb03, au111_Henk-jpcm04, au111_Mazarello-surfsci08, agau111_Nuber-prb11} and also at Au(110),~\cite{au110_Nuber-prb08, auagsb_Nagano-jpcm09, au110_Simon-prb10} through Li/W(110) and Li/Mo(110) overlayers,~\cite{liwmo110_Rotenberg-prl99} and
the Gd(0001) surface,~\cite{gd0001_Krupin-prb05} to a large number of metallic
surfaces and surface alloys related to Bi, Pb or Sb where the $5p$ and $6p$ orbitals show
a pronounced spin-orbit splitting.~\cite{bi111_Korotoev_prl04, bi100_Hofmann-prb05, sb111_Sugawara-prl06, bi100_Hirahara-prl06, pbag111_Pacile-prb06, biag111_Ast-prl07, bipbag111_Bihlmayer-prb07, bipbag111_Ast_prb07, pbsi111_Dil-prl08, bipbag111_Ast_prb08, pbag111_Hirahara-prb08, sbag111_Moreschini-prb09, bisbcu111_Moreschini-prb09, bipbsbag111_Gierz-prb10, bibatio3_Mirhosseini-prb10, bibatio3_Abdelouahed-prb10}
This huge interest is mainly triggered by potential spintronics applications in relation to
the Datta-Das transistor,~\cite{zutic04} the spin Hall effect~\cite{spinHall} and the
anomalous Hall effect.~\cite{AHE}

While accurate ab initio calculations satisfactorily account for most features of the measured
dispersion relations of metallic surface states, there is an obvious need to explain the RB effect
in terms of simple models containing a few, easily identifiable parameters. The simplest
effective Hamiltonian of a two-dimensional electron gas, subject to spin-orbit interaction (SOI),
includes in addition to the kinetic energy, $\varepsilon_0+\frac{\hbar^2 \mathbf{k}^2}{2m^\ast}$
($\mathbf{k}$ and $m^\ast$ being the wavevector and the effective mass of the electrons,
respectively), a Rashba term,~\cite{Bychkov-Rashba-jpcm84,PetersenHedegard-ss00}
\begin{equation}
H_{R}\left(  \mathbf{k}\right)=\alpha_{R}\left(  k_{x}\sigma_{y}-k_{y}\sigma_{x}\right)  \;,
\label{eq:RH-iso}
\end{equation}
where $\alpha_{R}$ is the so-called Rashba parameter and $\sigma_{i}$ ($i=x,y,z$) denote the Pauli
matrices.
The corresponding eigenvalues, $\varepsilon_{\pm}\left(  \mathbf{k}\right)
=\varepsilon_0 + \frac{\hbar^{2}\mathbf{k}^{2}}{2m^{\ast}}
\pm\alpha_{R}\,k$ ($k=\left\vert \mathbf{k}\right\vert$)  show an isotropic splitting
for $\mathbf{k} \ne 0$, and, at least for moderate values of $k$, they readily
can be fit to most experimental and ab initio dispersion relations.

Although not yet detected experimentally,~\cite{au110_Nuber-prb08}
 a RB splitting that is anisotropic in $k$-space is obvious
for the surface states at Au(110). The $C_{2v}$ point-group symmetry~\cite{Altmann-Herzig} of the system
not only implies an anisotropy of the effective mass, $m_{x}^{\ast}\neq m_{y}^{\ast}$,
but, as discussed in terms of $k \cdot p$ perturbation
theory,~\cite{Oguchi-jpcm09,au110_Simon-prb10}
it leads to a Rashba Hamiltonian containing
two independent Rashba parameters, $\alpha_{R}^1$ and $\alpha_{R}^2$, 
\begin{equation}
H_R\left(  \mathbf{k}\right)  =
\alpha_{R}^1\,k_{x}\sigma_{y} + \alpha_{R}^2\,k_{y}\sigma_{x} \;.
\label{eq:RH-c2v}%
\end{equation}
Fully relativistic ab initio calculations confirmed the existence of the anisotropic RB splitting at
 Au(110),~\cite{auagsb_Nagano-jpcm09} matching with a high accuracy to the eigenvalues of the
effective Hamiltonian in Eq.~(\ref{eq:RH-c2v}).~\cite{au110_Simon-prb10}

Even in case of high symmetry surfaces, i.e., having a point-group of $C_{3v}$ or $C_{4v}$, several studies~\cite{bi111_Korotoev_prl04, sb111_Sugawara-prl06, biag111_Ast-prl07, sbag111_Moreschini-prb09, bipbsbag111_Gierz-prb10, bibatio3_Mirhosseini-prb10} called the attention to an anisotropic RB splitting.
In Ref.~\onlinecite{bipbag111_Premper-prb07} the anisotropic RB effect at Bi/Ag(111) and Pb/Ag(111) surfaces
was reproduced by using a nearly-free electron model and explained due to in-plane structural
inversion asymmetry.
From the group-theoretical analysis in Ref.~\onlinecite{Oguchi-jpcm09} it is, however, clear that
under $C_{3v}$ and $C_{4v}$ point-group symmetry an effective $2\times2$ Hamiltonian
that is linear in the components of $\mathbf{k}$ must be of the form of Eq.~(\ref{eq:RH-iso}),
hence, it can not explain the observed anisotropy of the RB splitting. Thus, we conclude
that in these systems the anisotropic RB effect can be described by a Hamiltonian containing at least
third-order polynomials of $k_x$ and $k_y$. It should be noted that the second-order terms are
related to the kinetic energy (effective mass terms) that are irrelevant to the RB splitting.

To construct Rashba Hamiltonians up to third order in $k$,
in the present work we use $k \cdot p$ perturbation theory and group-theoretical methods different
from Ref.~\onlinecite{Oguchi-jpcm09}. Our analysis of the effective Hamiltonian
is closely related to that of Ref. \onlinecite{Fu-prl09}, where, for the case of $C_{3v}$
symmetry, the correct form
of $H({\bf k})$ is derived up to third order in $k$ and the corresponding band dispersion
was used to explain the hexagonal warping of the surface states' Fermi contour observed
experimentally in the topological insulator Bi$_2$Te$_3$.\cite{Chen-science09}

 We also perform relativistic ab initio calculation for the Bi/Ag(111) ordered alloy in
$(\sqrt{3}\times\sqrt{3})  R30^{\circ}$ superstructure and confirm that for higher values of $k$,
but still in the measured range, third-order terms
of the Rashba Hamiltonian are needed to reproduce the RB splitting and that these terms are of
the predicted form. Moreover, using explicit expressions of the third-order Rashba parameters within $k \cdot p$ perturbation
theory and calculated spectral densities at the Brillouin zone
center we are able to give a simple explanation why the Au(111) and the Bi/Ag(111) surface states
exhibit isotropic and anisotropic RB splitting, respectively.

\section{Perturbation theory and symmetry analysis}

Let \textbf{Q} be a high symmetry point of the surface Brillouin zone (SBZ), for which
a pair of spin-degenerate eigenstates exists on a nonmagnetic surface.
Due to time reversal symmetry, this is always the case if
$\textbf{Q} = -\textbf{Q} + \textbf{K}$ is satisfied where \textbf{K} is a
two-dimensional (2D) reciprocal-lattice vector.
Such points are the center of the SBZ ($\overline{\Gamma}$)
and some special points at the boundary of SBZ such as the $\overline{\rm X}$, $\overline{\rm Y}$
and $\overline{\rm S}$ points for a primitive rectangular lattice, the $\overline{\rm M}$ and $\overline{\rm X}$
points for a square lattice and the $\overline{\rm M}$ point for a hexagonal lattice.
As what follows, our investigations will concern solely this case termed as the proper Rashba effect.\cite{auagsb_Nagano-jpcm09}
As pointed out in Ref.~\onlinecite{auagsb_Nagano-jpcm09},
due to double-group symmetry there can happen degeneracy at points of the SBZ
that doesn't meet the above condition, like the $\overline{\rm K}$ point for
a hexagonal lattice (improper Rashba effect).

To describe the surface band around \textbf{Q} it is worth to label the corresponding Bloch states
by the wavenumber with respect to \textbf{Q}, $\psi_{\mathbf{Q+k}}$, and introduce a new
wavefunction, $\phi_{\mathbf{k}}$ as
\begin{equation}
\psi_{\mathbf{Q+k}}\left(  \mathbf{r}\right)  =e^{i\mathbf{kr}}\phi
_{\mathbf{k}}\left(  \mathbf{r}\right)  \;,
\label{eq:psik}
\end{equation}
with the boundary condition,
$\phi_{\mathbf{k}}(\mathbf{r+T})=e^{i\mathbf{QT}} \phi_{\mathbf{k}}(\mathbf{r})$,
where \textbf{T} is a 2D real-lattice vector.
Considering the Hamilton operator, $\mathcal{H}=\frac{\mathbf{p}^2}{2m}+V+\mathcal{H}_{SO}$, with
the crystal potential, $V$, and
$\mathcal{H}_{SO}$ denoting the spin-orbit interaction,
\begin{equation}
\mathcal{H}_{SO}=\frac{\hbar}{4m^{2}c^{2}}\left(
\boldsymbol{\nabla}V\times\mathbf{p}\right) \cdot \boldsymbol{\sigma} \:,
\end{equation}
the wavefunctions $\phi_{\mathbf{k}}$ satisfy the eigenvalue equation,
\begin{equation}
\left( \mathcal{H}_0({\bf k}) + \mathcal{H}_{SO}({\bf k}) \right) \phi_{\mathbf{k}}
= \varepsilon({\bf k}) \phi_{\mathbf{k}} \: ,
\end{equation}
with
\begin{equation}
\mathcal{H}_0({\bf k})=\frac{\left( \hbar {\bf k} + \mathbf{p} \right)^{2}}{2m}+V \: ,
\end{equation}
and
\begin{equation}
\mathcal{H}_{SO}(\mathbf{k}) =
\frac{\hbar}{4m^{2}c^{2}}\left(
\boldsymbol{\nabla}V\times  \left( \hbar \mathbf{k} + \mathbf{p}\right)
\right) \cdot \boldsymbol{\sigma} \: .
\label{HSO-k}
\end{equation}

Following the recipe used in Ref.~\onlinecite{au110_Simon-prb10}, in the first step we look for the
solution of the Schr\"odinger equation,
\begin{equation}
\mathcal{H}_0({\bf k}) \phi^0_{\mathbf{k}}
= \varepsilon_0({\bf k}) \phi^0_{\mathbf{k}} \: ,
\label{eq:SchE_0}
\end{equation}
which can be elucidated, e.g. in terms of $k \cdot p$ perturbation theory. Although such a calculation
provides with a deeper insight into the problem,\cite{au110_Simon-prb10}
in this section we just make use of the symmetry properties of the solutions, $\varepsilon_0({\bf k})$ and
$\phi^0_{\mathbf{k}}$. First we note that since $\mathcal{H}_0({\bf k})$ is independent
of the spin, the solutions of Eq.~(\ref{eq:SchE_0}) remain degenerate in spin-space.
Time reversal symmetry, $T{\mathcal H}_0({\bf k}) T^{-1}={\mathcal H}_0(-{\bf k})$ 
with $T \psi = \psi^\ast$, then immediately implies
\begin{eqnarray}
\varepsilon_0(-{\bf k})&=& \varepsilon_0({\bf k}) \: ,
\label{eq:e0-k} \\
\phi^0_{-\mathbf{k}} &=& \left( \phi^0_{\mathbf{k}} \right)^\ast \: ,
\label{eq:phi0-k}
\end{eqnarray}
where the phase of $\phi^0_{\mathbf{k}}$ has been fixed without loss of generality.
Clearly from Eq.~(\ref{eq:e0-k}), a polynomial form of $\varepsilon_0({\bf k})$ contains
just even powers: the first non-trivial (second-order) terms are obviously related to the effective masses.

We can draw further relations from point-group symmetry. Let $G_{\mathbf{Q}}$ be the
small group of {\bf Q}, i.e. $g {\bf Q} = {\bf Q} + {\bf K}$ for any $g\in G_{\mathbf{Q}}$
and {\bf K} denoting an appropriate reciprocal-lattice vector.
Using the standard definition for the action of a symmetry operation,
$(g \circ f)({\bf r}) = f(g^{-1}{\bf r})$, from the symmetry of the Hamilton operator,
$g \circ \mathcal{H}_0({\bf k}) = \mathcal{H}_0(g{\bf k})$, one easily can derive 
\begin{eqnarray}
\varepsilon_0(g{\bf k})&=& \varepsilon_0({\bf k}) \: ,
\label{eq:e0-gk} \\
g \circ \phi^0_{\mathbf{k}} &=& \phi^0_{g\mathbf{k}} \: .
\label{eq:phi0-gk}
\end{eqnarray}

In the second step, using $\mathcal{H}_{\rm SO}({\bf k})$ as perturbation and
$\phi^0_{\mathbf{k}} \, \chi_{s}$ with $\chi_{s}$ being spinor eigenfunctions
$\left( s = \pm \frac{{\footnotesize 1}}{{\footnotesize 2}} \right)$
as unperturbed wavefunctions, first-order degenerate
perturbation theory is applied. The Rashba Hamiltonian, $H_R({\bf k})$, is defined as the
corresponding $2 \times 2$ matrix,
\begin{equation}
H_R({\bf k}) = \boldsymbol{\alpha}({\bf k}) \cdot \boldsymbol{\sigma} \: ,
\label{eq:HRk}
\end{equation}
where
\begin{equation}
\boldsymbol{\alpha}(\mathbf{k})=\left\langle \phi^0_{\mathbf
{k}}\right\vert \frac{\hbar}{4m^{2}c^{2}}
\left(\boldsymbol{\nabla}V\times \left( \hbar \mathbf{k} + \mathbf{p} \right) \right)
\left\vert \phi^0_{\mathbf{k}}\right\rangle \: .
\label{eq:alphak}
\end{equation}

Our present goal is to derive the polynomial form of $\boldsymbol{\alpha}(\mathbf{k})$.
To this end we note two symmetry properties that can be obtained from Eqs.~(\ref{eq:phi0-k}) and
(\ref{eq:phi0-gk}):
\begin{equation}
\boldsymbol{\alpha}(-\mathbf{k})=-\boldsymbol{\alpha}(\mathbf{k}) \: ,
\label{eq.alpha-k}
\end{equation}
stating that $\alpha_i(k_x,k_y)$ can be expanded in terms of polynomials of odd power
and
\begin{equation}
\boldsymbol{\alpha}(g \, \mathbf{k})=det(g) \, g\boldsymbol{\alpha}(\mathbf{k}) \: ,
\label{eq:alpha-gk}
\end{equation}
where $det(g)=1$ for proper rotations and $det(g)=-1$ for improper rotations.
Eq.~(\ref{eq:alpha-gk}) is then used to set up linear equations for
the coefficients, $c_i^l$, of the $n^{th}$-order polynomials of $\alpha_i(k_x,k_y)= \sum_{l=1,\dots,n} c_i^l k_x^l k_y^{n-l}$ $(i=x,y,z)$. Solving this set of linear equations serves to
search for the vanishing coefficients, in principle, for any power $n$, hence,
to determine the form of $H_R({\bf k})$.

Another systematic way to obtain $H_R({\bf k})$ relies on
the observation that Eq.~(\ref{eq:alpha-gk}) can be used to formulate the invariance
of the Rashba Hamiltonian as
\begin{equation}
H_R({\bf k}) =  \boldsymbol{\alpha}(g \, \mathbf{k}) \cdot \left( det(g) \, g \boldsymbol{\sigma} \right)
 \, ,
\label{eq:HR-gk-invariance}
\end{equation}
also implying that $\boldsymbol{\sigma}$ transforms as an axial vector.
Sorting out the components of $\mathbf{k}$ and $\boldsymbol{\sigma}$ according to irreducible
representations of $G_{\bf Q}$, their direct products can again be decomposed into irreducible
representations.
Eq.~(\ref{eq:HR-gk-invariance}) states that only the total symmetric irreducible representations
from this decomposition can contribute to $H_R({\bf k})$. From the corresponding tables
of the point groups~\cite{Altmann-Herzig} one can easily construct the
possible terms entering $H_R({\bf k})$ according to increasing powers of $k_x$ and $k_y$.
In the Appendix this procedure is illustrated for the simple case of point group $C_{2v}$.
As one of the main result of this work, in Table \ref{table1}
we list the possible terms up to third order in $k$
that can enter $H_R({\bf k})$ for different point groups relevant to surfaces of crystals.

\onecolumngrid
\ \vskip -10pt
\hrulefill
\begin{center}
\begin{table}[htb]
\begin{tabular}{|l|l|l|l|l|l|l|}
\hline
\multicolumn{1}{|c|}{}&
\multicolumn{1}{c}{} &
\multicolumn{1}{|c|}{}&
\multicolumn{1}{c}{}&
\multicolumn{1}{|c|}{}&
\multicolumn{1}{c}{}&
\multicolumn{1}{|c|}{} \\ [-1.5ex]
\multicolumn{1}{|c|}{$C_{hx}$} &
\multicolumn{1}{c}  {$C_{2}$} &
\multicolumn{1}{|c|}{$C_{3}$} &
\multicolumn{1}{c}  {$C_{4}$} &
\multicolumn{1}{|c|}{$C_{2v}$} &
\multicolumn{1}{c}  {$C_{3v}$} &
\multicolumn{1}{|c|}{$C_{4v}$} \\ [0.5ex]
\hline \hline  &&&&&&  \\ [-1.5ex]
$k_{x}\sigma_{y}\,,$ &
$k_{x}\sigma_{x}\,,$ &
$k_{x}\sigma_{x}+k_{y}\sigma_{y}\,,$ &
$k_{x}\sigma_{x}+k_{y}\sigma_{y}\,,$ &
$k_{x}\sigma_{y}\,,$ &
$k_{x}\sigma_{y}-k_{y}\sigma_{x}$  &
$k_{x}\sigma_{y}-k_{y}\sigma_{x}$ \\ [0.5ex]
$k_{y}\sigma_{x}$\,, &
$k_{x}\sigma_{y}\,,$ &
$k_{x}\sigma_{y}-k_{y}\sigma_{x}$ &
$k_{x}\sigma_{y}-k_{y}\sigma_{x}$ &
$k_{y}\sigma_{x}$ && \\ [0.5ex]
$k_y\sigma_{z}$  &  $k_{y}\sigma_{x}\,,$ &&&&& \\ [0.5ex]
  &  $k_{y}\sigma_{y}$   &&&&& \\  [0.5ex]
\hline  &&&&&&  \\ [-1.5ex]
$k_{x}^{3}\sigma_{y}\,,$ &
$k_{x}^{3}\sigma_{x}\,,$ &
$\left(k_{x}^{3}+k_{x}k_{y}^{2}\right)\sigma_{x}+$ &
$k_{x}^{3}\sigma_{x}+k_{y}^{3}\sigma_{y}\,,$  &
$k_{x}^{3}\sigma_{y}\,,$ &
$\left(k_{x}^{3}+k_{x}k_{y}^{2}\right)\sigma_{y}-$  &
$k_{x}^{3}\sigma_{y}-k_{y}^{3}\sigma_{x}\,,$ \\ [0.5ex]
$k_{x}^{2}k_{y}\sigma_{x}\,,$ &
$k_{x}^{3}\sigma_{y}\,,$ &
$\left(k_{x}^{2}k_{y}+k_{y}^{3}\right) \sigma_{y}\,,$ &
$k_{x}^{3}\sigma_{y}-k_{y}^{3}\sigma_{x}\,,$ &
$k_{x}^{2}k_{y}\sigma_{x}\,,$   &
$\left(k_{x}^{2}k_{y}+k_{y}^{3}\right) \sigma_{x}\,,$ &
$k_{x}^{2}k_y\sigma_{x}-k_xk_{y}^{2}\sigma_{y}$ \\ [0.5ex]
$k_{x}^{2}k_{y}\sigma_{z}\,,$ &
$k_{x}^{2}k_{y}\sigma_{x}\,,$ &
$\left(k_{x}^{3}+k_{x}k_{y}^{2}\right)\sigma_{y}-$ &
$k_{x}^{2}k_{y}\sigma_{x}-k_{x}k_{y}^{2}\sigma_{y}\,,$ &
$k_xk_{y}^{2}\sigma_{y}\,,$ &
$\left(k_{x}^{3}-3k_{x}k_{y}^{2}\right) \sigma_{z}$  & \\ [0.5ex]
$k_{x}k_{y}^{2}\sigma_{y}\,,$ &
$k_{x}^{2}k_{y}\sigma_{y}\,,$ &
$\left(k_{x}^{2}k_{y}+k_{y}^{3}\right) \sigma_{x}\,,$ &
$k_{x}k_{y}^{2}\sigma_{x}+k_{x}^{2}k_{y}\sigma_{y}$ &
$k_{y}^{3}\sigma_{x}$ &&  \\ [0.5ex]
$k_{y}^{3}\sigma_{x}\,,$ &
$k_{x}k_{y}^{2}\sigma_{x}\,,$ &
$\left(  k_{x}^{3}-3k_{x}k_{y}^{2}\right) \sigma_{z}\,,$  &&&&  \\ [0.5ex]
$k_{y}^{3}\sigma_{z}$ &
$k_{x}k_{y}^{2}\sigma_{y}\,,$  &
$\left( k_{y}^{3}-3k_{x}^{2}k_{y}\right) \sigma_{z}$  &&&&  \\ [0.5ex]
& $k_{y}^{3}\sigma_{x}\,,$ &&&&&  \\ [0.5ex]
& $k_{y}^{3}\sigma_{y}$ &&&&&  \\ [0.5ex]
\hline
\end{tabular}
\caption{Possible terms of the Rashba Hamiltonian for different point groups (first row)
containing first-order (second row) and third-order (third row) polynomials of $k_x$ and $k_y$.
}
\label{table1}
\end{table}
\end{center}

\vskip -30pt
\hrulefill
\vskip 10pt
\twocolumngrid

Finally in this section, we comment on the method used in Ref.~\onlinecite{Oguchi-jpcm09}.
In this work a Hamiltonian including SOI but excluding all $k$-dependent terms was
considered as the unperturbed system and the twofold degenerate solutions, $\phi_1$ and $\phi_2$,
corresponding to the wavenumber {\bf Q} as the unperturbed solutions. The perturbation was therefore taken
as $\mathcal{H}^\prime(\mathbf{k})=\frac{\hbar}{m} \mathbf{k} \mathbf{p} +
\frac{\hbar^2}{4m^{2}c^{2}}\left( \boldsymbol{\nabla}V\times \mathbf{k}
\right) \cdot \boldsymbol{\sigma}$ and, similar to our strategy, first-order degenerate
perturbation theory was applied.
The form of the effective Rashba Hamiltonian, $H_{ij}^{\prime}(\mathbf{k})=\langle \phi_i \! \mid \!
\mathcal{H}^\prime(\mathbf{k}) \! \mid \! \phi_j \rangle$ ($i,j=1,2$), is then determined via the invariance conditions,
\begin{equation}
H^{\prime}(g\mathbf{k})=D\left(  g \right)  \,H^{\prime}%
(\mathbf{k})\,D\left( g \right)  ^{-1},%
\label{eq:Hg-doublegroup}
\end{equation}
where $D\left(  g\right)$ is a $2\times2$ unitary double-point group representation of $g$.
In case of Abelian point groups ($C_{hx}$, $C_{2}$, $C_{3}$ and $C_{4}$), the degenerate
states form time-reversed pairs and
$D(g)$ can simply be set up from the characters of the corresponding one-dimensional
irreducible representations.
Following from the definition of $H^{\prime}(\mathbf{k})$,
in Ref.~\onlinecite{Oguchi-jpcm09} the first-order Rashba Hamiltonians were obtained for the groups $C_{hx}$, $C_{2v}$, $C_{3v}$ and $C_{4v}$.
It is, however, straightforward to show that, when applied to the Hamiltonian (\ref{eq:HRk}),
Eq.~(\ref{eq:Hg-doublegroup}) is equivalent with condition (\ref{eq:HR-gk-invariance}),\cite{Vajna-2010}
hence, using double-group representations leads to the same results as listed in Table~\ref{table1}.

\section{Third-order Rashba splitting at Bi/Ag(111)}

By using the relativistic Screened Korringa-Kohn-Rostoker (KKR) method~\cite{SKKR-book}
we performed calculations for the $(\sqrt{3}\times\sqrt{3})R30^{\circ}$ ordered surface alloy Bi/Ag(111)
to obtain a quantitative verification of our prediction of a third-order Rashba Hamiltonian.
A 2D lattice constant of 2.892 \AA \ related to fcc Ag bulk and, according to
geometry optimization we performed in terms of the VASP method\cite{VASP} and also
in agreement to previous LAPW
calculations,\cite{bipbag111_Bihlmayer-prb07} an outward buckling of 36~\% (0.85 \AA)
for the Bi atoms were considered.
The local spin-density approximation as parametrized by Vosko \emph{et al. }\cite{voskoCJP80} was applied,
the effective potentials and fields were treated within the atomic sphere approximation with an angular momentum cut--off of $\ell_{max}=2$.  The energy integrations were performed by sampling 12 points on a semi-circular path in the upper complex semi-plane and for the necessary {\it k}-integrations we selected 36 {\em k}-points in the irreducible segment of the surface Brillouin zone.

\begin{figure}[h!]
\begin{center}
\includegraphics[width=0.40\textwidth,bb=95 50 325 270,clip]{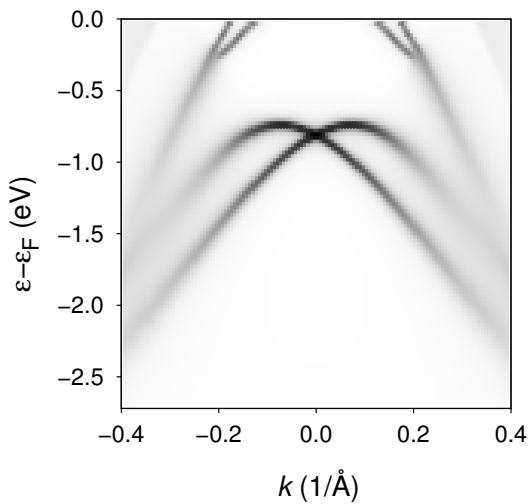}
\vskip -0.3cm
\caption{Calculated dispersion relations of the occupied surface states of Bi/Ag(111) along the $\overline{\Gamma}-\overline{M}$ direction of the SBZ.}
\label{fig:disp}
\end{center}
\end{figure}

The calculated dispersion relation of the Bi surface states below the Fermi level is shown
 in Fig.~\ref{fig:disp} along the $\overline{\Gamma}-\overline{M}$ direction of the SBZ.
The maxima of the Rashba-split Bi $sp_z$ band are shifted from the $\overline{\Gamma}$ point by
$k_0 =0.1$ \AA$^{-1}$ and, using a parabolic fit around the maxima, we obtained an effective mass of
$m^{\star }=-0.36 \, m_{e}$. These values are in good agreement with experimental data,
$k_0 =0.13$ \AA$^{-1}$ and $m^{\star }=-0.35 \, m_{e}$.~\cite{biag111_Ast-prl07}
It should, however, be mentioned that the calculated surface bands are shifted downwards by about
0.5 eV as compared to experiment, most probably, due to the angular momentum cutoff and
to the atomic sphere approximation used in the calculations.
The Bi $p_x p_y$ surface bands, shifted upwards due to crystal field splitting
and to spin-orbit coupling, can also be clearly seen in Fig.~\ref{fig:disp}.

In case of $C_{3v}$ symmetry the effective Rashba Hamiltonian can be written up to third order in $k$ as
(see Table \ref{table1}),
\begin{equation}
H_{R}(\underline{k})=\left(\alpha_1+\alpha_{3}^{1}\,k^{2}\right)\left(  k_{x}\sigma
_{y}-k_{y}\sigma_{x}\right)  + a_{3}^{2} k^{3} \cos 3\varphi \, \sigma_{z} \, ,
\label{eq:RH-c3v}
\end{equation}
with the polar coordinate $\varphi={\rm arccos}(k_x/k)$.
Note that the $x$ axis was chosen along the  $\overline{\Gamma}-\overline{\rm K}$ direction of the SBZ.
Obviously, there are two kinds of third-order contributions to the Hamiltonian (\ref{eq:RH-c3v}):
an isotropic one with coefficient $\alpha_{3}^{1}$ and an anisotropic one with the
coefficient $\alpha_{3}^{2}$.
The square of the splitting of the eigenvalues, $\Delta \varepsilon(\mathbf{k}) =
(\varepsilon_+(\mathbf{k}) - \varepsilon_-(\mathbf{k}))/2$, can then be expressed as
\begin{equation}
\Delta \varepsilon(\mathbf{k})^2 =
\left(  \alpha_{1} k+\alpha_{3}^{1}\,k^{3}\right)^{2}+\left(
\alpha_{3}^{2}\right)^{2}\,k^{6}\,\cos^{2}3\varphi  \: .
\label{eq:de2}
\end{equation}

\begin{figure}[h!]
\begin{center}
\includegraphics[width=0.45\textwidth,bb=13 10 180 130,clip]{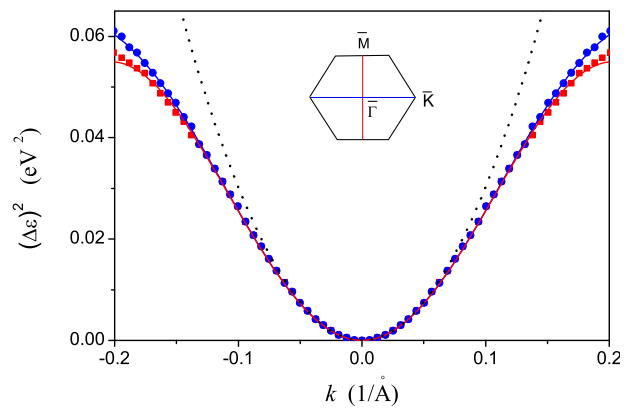}
\caption{Square of the calculated splitting, $\Delta \varepsilon(\mathbf{k}) =
(\varepsilon_+(\mathbf{k}) - \varepsilon_-(\mathbf{k}))/2$, of the occupied surface states of Bi/Ag(111).
Squares: $\overline{\Gamma}-\overline{\rm K}$ direction, circles: $\overline{\Gamma}-\overline{\rm M}$
direction, see the sketch of the SBZ in the inset. Dotted and solid lines display first-order and
third-order fits as described in the text.}
\label{fig:ediff}
\end{center}
\end{figure}

In Fig. \ref{fig:ediff} we plotted $\Delta \varepsilon(\mathbf{k})^2$ along the
$\overline{\Gamma}-\overline{\rm K}$ and the $\overline{\Gamma}-\overline{\rm M}$
directions, together with different fitting functions related to Eq.~(\ref{eq:de2}).
It can be seen that a parabolic fit (dots), $\alpha_1 k^2$ with $\alpha_1= 1.74 \, {\rm eV} \, {\rm \AA}$,
applies well to the two curves only for about $k < 0.07$ \AA$^{-1}$. Up to $k \sim 0.13$ \AA$^{-1}$ \
the two curves still coincide, however, the isotropic third-order contribution
is needed for a good fit: here we used a fitting function
$\left(  \alpha_{1} k+\alpha_{3}^{1}\,k^{3}\right)^{2}$ with
the same value for $ \alpha_{1}$ as before and $\alpha_3^1 = -14.2 \, {\rm eV} \, {\rm \AA}^3$.
For wavenumbers $k >  0.13$ \AA$^{-1}$, the anisotropy of the RB splitting becomes apparent:
along $\overline{\Gamma}-\overline{\rm M}$ ($\phi = \pi/2$) the previous fit applies, while
along $\overline{\Gamma}-\overline{\rm K}$ ($\phi = 0$) the fitting function had to be
extended by the anisotropic term, $\left(\alpha_{3}^{2}\right)^{2}\,k^{6}$ with
$\alpha_3^2 = 9.4 \, {\rm eV} \, {\rm \AA}^3$. Our numerical results thus clearly support
the appearance of third-order terms for the surface states of Bi/Ag(111) consistent with the functional form
as derived from group-theoretical methods.

\section{Comparison of the Rashba effect at Au(111) and Bi/Ag(111)}
It is well-known from experiments and ab inito calculations,\cite{au111_LaShell-prl96,auag111_Nicolay-prb01,au111_Henk-prb03,au111_Henk-jpcm04,au111_Mazarello-surfsci08}
that the Au(111) $L$-gap surface states show a highly isotropic
(first-order) Rashba splitting.
Since both systems, Au(111) and Bi/Ag(111), exhibit $C_{3v}$ symmetry,
the question naturally arises why there is a remarkable difference concerning third-order
RB splitting. In order to find, at least, a qualitative understanding of the problem we
extended the $k \cdot p$ perturbation calculations presented
in Ref.~\onlinecite{au110_Simon-prb10} for the case of $C_{2v}$ symmetry
to $C_{3v}$ symmetry and found the following
expressions for the third-order coefficients in Eq.~(\ref{eq:RH-c3v}),
\begin{equation}
\alpha _{3}^{1} =\frac{\hbar^{4}}{4m^{4}c^{2}}
{\textstyle\sum\limits_{n,m}}\frac{\langle\phi_{0}|\,p_{x}\,|\,\phi_{n}^{+}\rangle\langle
\phi_{n}^{+}|\partial_{z}V|\phi_{m}^{+}\rangle \langle \phi_{m}^{+}|p_{x}|\phi
_{0}\rangle }{(\varepsilon_{0}-\varepsilon_{n}^E)(\varepsilon_{0}-\varepsilon_{m}^E)  },
\label{eq:a31}
\end{equation}
and
\begin{equation}
\alpha _{3}^{2} =\frac{\hbar^{4}}{4m^{4}c^{2}}
{\textstyle\sum\limits_{n,m}}\frac{\langle\phi_{0}|\,p_{x}\,|\,\phi_{n}^{+}\rangle\langle
\phi_{n}^{+}|\partial_{x}V|\phi_{m}^{-}\rangle \langle \phi_{m}^{-}|p_{x}|\phi
_{0}\rangle }{(\varepsilon_{0}-\varepsilon_{n}^E)(\varepsilon_{0}-\varepsilon_{m}^E)  }.
\label{eq:a32}
\end{equation}
From the above formulas it turns out that third-order corrections to the effective Hamiltonian
arise from an admixture between the surface state, $\phi_0$ of  $sp_z$ orbital character
at energy $\varepsilon_0$, and
those corresponding to the two-dimensional irreducible representation, $E$, $\phi_{n}^{\pm}$
with $p_x \pm i p_y$ character, at energy $\varepsilon_n^E$. Note that all these states are
eigenstates of the Hamiltonian, $\mathcal{H}=\frac{p^2}{2m}+V$, at the center of the surface band {\bf Q}.
It is remarkable that, similar to the isotropic first-order Rashba parameter, the strength of the
isotropic contribution, $\alpha _{3}^{1}$, depends on the
partial derivative of the crystal potential normal to the surface, $\partial_z V$,
while the coefficient for the anisotropic term, $\alpha _{3}^{2}$, is related to the in-plane
gradient of the potential, $\partial_x V$.

\begin{figure}[h!]
\begin{center}
\includegraphics[width=0.45\textwidth,bb=16 16 330 420,clip]{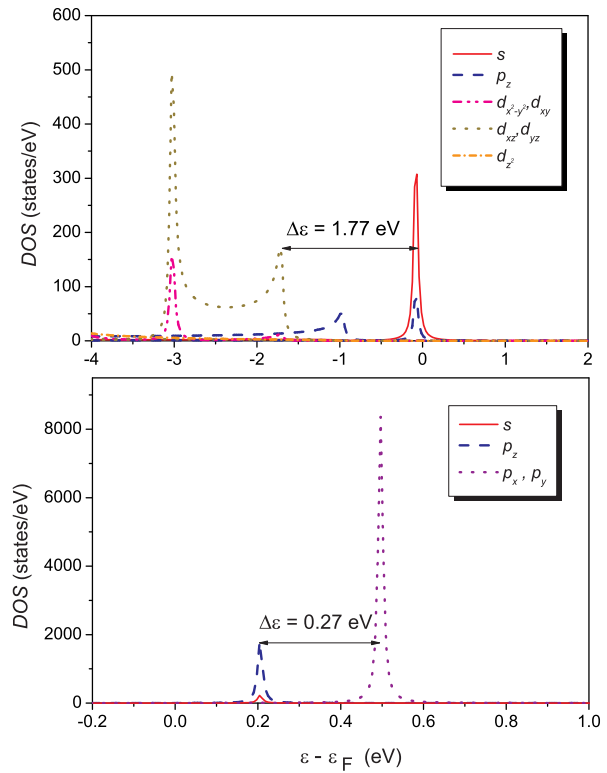}
\caption{Calculated orbital projected partial densities of states at the $\overline{\Gamma}$ point around
the surface states of Au(111) (upper panel) and Bi/Ag(111) (lower panel). }
\label{fig:band}
\end{center}
\end{figure}

In Fig.~\ref{fig:band} we plotted the scalar-relativistic, orbital projected partial densities of states
(Bloch spectral functions) at the $\overline{\Gamma}$ point in an appropriate energy window
around the surface states of Au(111) and Bi/Ag(111).
From the upper panel it can be seen that in case of Au(111) the edge of the bulk subband of $E$
symmetry closest to the surface state is by $\Delta \varepsilon=1.77$~eV below $\varepsilon_0$.
The situation is entirely different for Bi/Ag(111), since the Bi $p_x p_y$ states of $E$ symmetry
are separated from the $sp_z$ state by only $\Delta \varepsilon=0.27$~eV.
Since these are the characteristic energy differences that enter
the denominators in Eqs.~(\ref{eq:a31}) and (\ref{eq:a32}), a difference of at least two orders
in magnitude can indeed be estimated concerning the third-order RB effect.
Most probably, the actual values of the matrixelements of $p_x$ and $\partial_{x,z}V$
even further strengthen this difference.

\section{Conclusions}
Based on $ k \cdot p$ perturbation theory including spin-orbit interaction we gave a
suitable definition to an effective Hamiltonian, Eqs.~(\ref{eq:HRk}) and (\ref{eq:alphak}),
describing the Rashba-Bychkov splitting on metallic surfaces.  Due to time reversal
and point-group symmetry we showed how to obtain the most general forms for the effective Hamiltonian,
and derived them up to third order in $k$ for point groups compatible with surfaces of real crystals.
Since the effective Hamiltonan (\ref{eq:HRk}) applies to a couple of non-interacting two-band
models, the expressions listed in Table~\ref{table1} can be used in quite a general sense.

Using the relativistic Screened Korringa-Kohn-Rostoker method, we demonstrated that the Rashba
splitting of the Bi $sp_z$ surface band of the ordered surface alloy Bi/Ag(111) can not
be satisfactorily described in terms of a first-order isotropic Rashba Hamiltonian.
Moreover, we showed that the strong third-order contribution is subject to an anisotropy
consistent with the dispersion relation deduced from our symmetry analysis.

We also derived explicit formulas for the third-order anisotropy parameters and established that
the isotropic and anisotropic contributions are related to the normal-to-plane and the in-plane
gradients of the crystal potential, respectively. Comparing the energy separation of relevant
 orbital projected bands for Au(111) and Bi/Ag(111), the derived expressions were
useful to give a qualitative
understanding of the different nature of Rashba-Bychkov splitting in these two systems.

\acknowledgments
The authors appreciate stimulating discussions with Gergely Zarand.
Financial support was provided by the Hungarian Research
Foundation (contract no. OTKA K77771, K84078 and PD83353) and by
the New Sz\'echenyi Plan of Hungary (Project ID: T\'AMOP-4.2.1/B-09/1/KMR-2010-0002).
K.P. kindly acknowledges support of the Bolyai Grant.

\bigskip
\appendix*
\section{Rashba Hamiltonians for point group $\boldsymbol{C_{2v}}$}
By using the direct products of irreducible representations, in this Appendix
we give an example for the polynomial forms of a $2 \times 2$ effective Hamiltonian for the
point-group $C_{2v}$. Let us denote the elements of the group by $E:\{x,y,z\}$,
$C_2:\{-x,-y,-z\}$, $S_x:\{-x,y,z\}$ and $S_y:\{x,-y,z\}$. The group has four one-dimensional
irreducible representations: $A_{1}$, $A_{2}$, $B_{1}$, $B_{2}$ with the character table,\cite{Altmann-Herzig}
\[%
\begin{tabular}
[c]{c|cccc}
& $E$ & $C_{2}$ & $S_{x}$ & $S_{y}$\\\hline
$A_{1}$ & $1$ & $1$ & $1$ & $1$\\
$A_{2}$ & $1$ & $1$ & $-1$ & $-1$\\
$B_{1}$ & $1$ & $-1$ & $-1$ & $1$\\
$B_{2}$ & $1$ & $-1$ & $1$ & $-1$%
\end{tabular}  \; .
\]
Making use that the {\bf k} and $\boldsymbol{\sigma}$ transform as polar and axial vectors,
respectively, a comparison with the character table lets us to sort out the components
of this vectors according to irreducible representations:  $ B_{1}:k_{x},\sigma_{y}$,
$B_{2}:k_{y},\sigma_{x}$  and $A_{2}:\sigma_{z}$.

From the table of direct products,
\[%
\begin{tabular}
[c]{c|cccc}
          & $A_{1}$ & $A_{2}$ & $B_{1}$ & $B_{2}$\\\hline
  $A_{1}$ & $A_{1}$ & $A_{2}$ & $B_{1}$ & $B_{2}$\\
  $A_{2}$ &         & $A_{1}$ & $B_{2}$ & $B_{1}$\\
  $B_{1}$ &         &         & $A_{1}$ & $A_{2}$\\
  $B_{2}$ &         &         &         & $A_{1}$
\end{tabular}  \; ,
\]
it is easy to find that the only combinations that are first order in $k_x$ and $k_y$
and correspond to the $A_1$ irreducible representations are $k_x \sigma_y$ and $k_y \sigma_x$,
therefore, the first-order Rashba Hamiltonian can be written in the form of Eq.~(\ref{eq:RH-c2v}).

The second-order polynomials of $k_x$ and $k_y$ can be sorted according to irreducible
representations as follows: $ A_{1}:k_{x}^2, k_{y}^2 $ and $A_{2}:k_{x} k_y$.
It should be noted that this implies the form of $\frac{\hbar^2}{2m_x^\ast} k_{x}^2
+ \frac{\hbar^2}{2m_y^\ast} k_{y}^2$ for the effective mass term.
The third-order polynomials of $k_x$ and $k_y$ can then be classified as
$B_{1}:k_{x}^3, k_x k_{y}^2$ and $B_{2}:k_{y}^3, k_y k_{x}^2 $.
Taking direct products with $\sigma_i$ of $A_1$ symmetry leads to the possible
third-order contributions to the Rashba Hamiltonian:
$k_{x}^{3}\sigma_{y}$, $k_{x}^{2}k_{y}\sigma_{x}$, $k_{x}k_{y}^{2}\sigma_{y}$
and $k_{y}^{3}\sigma_{x}$, i.e.
\begin{equation}
H_{R}^{3}(\mathbf{k})=\alpha_{3}^{1}\,k_{x}^{3}\sigma_{y}%
+\alpha_{3}^{2}\,k_{x}^{2}k_{y}\sigma_{x}+\alpha_{3}^{3}\,k_{x}k_{y}^{2}%
\sigma_{y}+\alpha_{3}^{4}\,k_{y}^{3}\sigma_{x}.
\end{equation}

\bibliographystyle{myst}


\end{document}